\documentclass[aps,preprint]{revtex4}
\usepackage{graphicx}
\begin{document}
\title{\Large \bf Superradiant Instability of Five-Dimensional
Rotating Charged  AdS Black Holes }
\author{\large Alikram N. Aliev and \"{O}zg\"{u}r Delice}
\address{Feza G\"ursey Institute, P. K. 6  \c Cengelk\" oy, 34684 Istanbul, Turkey}
\date{\today}

\begin{abstract}

We study the instability of small AdS black holes with two independent rotation parameters in minimal five-dimensional  gauged supergravity to massless scalar perturbations. We analytically solve the Klein-Gordon equation for low-frequency perturbations in two regions  of the spacetime of these black holes: namely, in the region close to the horizon  and in the far-region. By matching the solutions in an intermediate region, we calculate the frequency spectrum of quasinormal modes. We show that in the regime of superradiance only the modes of {\it even} orbital quantum number undergo negative damping, resulting in exponential growth of the amplitude. That is,  the black holes become unstable  to these modes. Meanwhile, the modes of  {\it odd} orbital quantum number do not undergo any damping, oscillating with frequency-shifts. This is in contrast with  the case of four-dimensional small Kerr-AdS black holes which exhibit the instability  to all modes of scalar perturbations in the regime of superradiance.

\end{abstract}


\maketitle

\section{Introduction}

Nowadays the classical theory of black holes in four-dimensional asymptotically flat spacetime  is thought of as elegant and well understood. General relativity provides  a unique family of exact solutions for stationary black holes which, in the most general case,
involves only three physical parameters: the  mass, angular momentum and the electric charge. Since classically the stationary black holes are ``dead" objects, it is of crucial importance to explore their characteristic responses to external perturbations of different sorts.  {\it Superradiance}  is one of such responses, namely, it is a phenomenon of amplification of scalar, electromagnetic and gravitational waves  scattered by a rotating  black hole.

Though the phenomenon  of superradiance, as a  Klein-paradox state of non-gravitational quantum systems, has been known for a long time (see \cite{mano} and references therein),  Zel'dovich was  the first to suggest the idea of supperradiant amplification of waves when scattering by the rotating black hole \cite{zeldovich}.  In order to argue the idea he had explored a heuristic model of the  scattering of a wave  by a rotating and absorbing cylinder. It turned out that when  the wave spectrum contains the frequency $ \omega $ fulfilling the condition  $ \omega < m \Omega $, where $ m $ is the azimuthal number or magnetic quantum number of the wave and $ \Omega $ is the angular velocity of the cylinder, the reflection of the wave occurs with amplification. In other words, the rotating cylinder effectively  acts as an  amplifier, transmitting its rotational energy to the reflected wave. Zel'dovich concluded that a similar phenomenon must occur with rotating black holes as well, where the horizon plays the role of an absorber. Similar arguments showing that certain modes of scalar waves must be amplified by a Kerr black hole were also given in \cite{misner}. A complete theory of the superradiance in the Kerr metric was developed by Starobinsky in \cite{starobinsky1}. (See also Ref.\cite{starobinsky2}). The appearance of the superradiance in string microscopic models of rotating black holes was studied in a recent paper \cite{dias1}.

Physically, the superradiant scattering is a  process of  stimulated radiation which emerges due to the excitations of  negative energy modes in the ergosphere of the  black hole. It is a wave analogue of the Penrose  process \cite{penrose}, in which a particle entering the ergosphere decays into two particles, one of which has a negative  energy relative to infinity and is absorbed by the black hole. This renders the other particle to leave the ergosphere with greater energy than the initial one, thereby extracting the rotational energy from the black hole. In a quantum-mechanical picture, the superradiance is of stimulated emission of quanta which  must be accompanied by their spontaneous emission as well \cite{zeldovich}. The spontaneous superradiance  arises due to quantum instability of the vacuum in the  Kerr metric, leading to a pair production of particles. When leaving the ergosphere these particles  carry positive energy and angular momentum from the black hole to infinity, whereas inside the ergosphere they form negative energy and angular momentum flows into the black hole \cite{unruh}.

The phenomenon of superradiance, after all,  has  a deep conceptual significance for understanding the stability properties of the black holes. As early as 1971 Zel'dovich \cite{zeldovich} noted that placing a reflecting mirror (a resonator) around a rotating black hole would result in re-amplification of superradiant modes and eventually the system would develop instability. The effect of the instability was later studied  in \cite{press1} and the system is now known as  a ``black hole bomb." This study has also created the motivation  to answer  general questions  on the stability of rotating black holes against small external perturbations. Using analytical and numerical methods, it has been shown that the Kerr black holes are stable  to massless scalar, electromagnetic, and gravitational perturbations \cite{press2}. However, the situation  turned out to be  different for perturbing massive bosonic fields. As is known,  classical  particles of energy $ E $  and mass $ m $, obeying the condition $ E < m  $,  perform a finite motion in the gravitational potential of the black hole. From quantum-mechanical point of view there  exists a certain probability for tunneling  such  particles through the potential barrier into the black hole. In consequence of this, the bound states of the particles inside the potential well must  become {\it quasistationary} or {\it quasinormal} (see, for instance \cite{ag1} and references therein). Similarly, for fields with mass $ \mu $, the wave of frequency $ \omega < \mu  $  can be thought of as a  ``bound particle" and therefore  must undergo repetitive reflections between the  potential well  and the horizon. In the regime of superradiance,  this will cause exponential growth of the number of particles in the quasinormal states, developing the instability \cite{damour, zouros, detweiler, dolan}. Thus, for massive bosonic fields the potential barrier of the black hole plays the role of a mirror in the heuristic model of the black hole bomb. There are also alternative models where a reflecting mirror leading to the instability arises due to an extra dimension which, from a Kaluza-Klein point of view, acts as a massive term (see for instance, Ref. \cite{dias2}).

In recent years, the question of the stability of black holes to external perturbations has been the subject of extensive studies in four and higher-dimensional spacetimes with a cosmological  constant. In particular, analytical and numerical works have revealed the perturbative stability of nonrotating  black holes in de Sitter or anti-de Sitter (dS/AdS) spacetimes  of various dimensions \cite{kodama1, konoplya}.  Though the similar general analysis concerning the stability of rotating black holes in the cosmological  spacetimes  still remains an open question,  significant progress has been achieved in understanding their superradiant instability \cite{hreall, cardoso1, cardoso2, cardoso3}.  The causal structure of the AdS spacetime  shows  that spatial infinity in it  corresponds to a finite region with a timelike boundary. Because of this property, the spacetime exhibits a ``box-like" behavior, ensuring the repetitive reflections of  massless bosonic waves  between spatial infinity and a Kerr-AdS black hole. The authors of work \cite{hreall} have shown that the Kerr-AdS black hole in five dimensions admits a corotating Killing vector which remains timelike everywhere outside the horizon, provided that the angular velocity of the boundary Einstein space does not exceed the speed of light. This means that there is no way to extract energy  from the black hole. However, these authors have also given simple arguments showing that for over-rotating Kerr-AdS black holes whose  typical size  is constrained  to $ r_{+} < l $, where $ l $ is a length scale determined by the negative cosmological constant, the superradiant instability may  occur. That is, the small Kerr-AdS black holes may become unstable against external perturbations. The idea was further developed in \cite{cardoso1, cardoso2}. In particular,
it was found that there must exist a  critical radius for the location of the mirror in the black hole bomb model. Below this radius the superradiant condition is violated and the system  becomes classically stable.  Extending this fact to the  case of the small Kerr-AdS black holes in four dimensions, the authors proved that the black holes indeed exhibit a superradiant instability to massless scalar perturbations. Later on, it was  shown that the small Kerr-AdS black holes are also unstable to gravitational perturbations \cite{cardoso3}.  In a recent work \cite{kodama2}, it was argued that the  instability properties of  the Kerr-AdS black holes  to gravitational perturbations are equivalent to those against massless scalar perturbations.

The main purpose of the present  paper is to address the superradiant instability of small rotating charged AdS black holes with two independent rotation parameters in minimal five-dimensional gauged supergravity. In Sec.II we discuss  some  properties of the spacetime metric given in the Boyer-Lindquist coordinates which are rotating at spatial  infinity. In particular, we define a  corotating Killing vector and calculate the angular velocities of the horizon as well as its electrostatic potential. We also discuss the ``hidden" symmetries of the metric and demonstrate the separability of the Hamilton-Jacobi equation for massive charged particles. Section III is devoted to the study of the Klein-Gordon equation. We show that it is  completely separable for massive charged particles and present the decoupled radial and angular equations in the most compact form. In Sec.IV we consider the near-horizon behavior of the radial equation and find the threshold frequency for the superradiance. In Sec.V we examine the instability  of  the small AdS black holes to low-frequency scalar perturbations.  Here we construct  the solution of the radial equation in the region close to the horizon and in the far-region. By matching these solutions in an intermediate region, we obtain the frequency spectrum for the quasinormal modes. We show that in the regime of superradiance the black hole exhibits instability  to ``selective" modes of the perturbations: Namely, only the modes of even orbital quantum number $ \ell $  exponentially grow with time. We also show that the modes of odd  $ \ell $ do not exhibit any damping, but oscillate with  frequency-shifts. In the Appendix we study the angular equation for AdS modified spheroidal harmonics in five dimensions.

\section{The metric and its properties}

The general  metric for  rotating charged AdS black holes in the bosonic sector of minimal supergravity theory in five dimensions was recently found in \cite{cclp}. The theory is described by  the action
\begin{eqnarray}
S&=& \int d^5x \sqrt{-g} \left(R+\frac{12}{l^2} -\frac{1}{4}\, F_{\alpha \beta}F^{\alpha \beta}
+\frac{1}{12\sqrt{3}}\,\epsilon^{\mu\nu \alpha\beta\lambda}F_{\mu\nu} F_{\alpha\beta}A_{\lambda}\right)\,,
\label{5sugraaction}
\end{eqnarray}
leading to the coupled  Einstein-Maxwell-Chern-Simons field equations
\begin{equation}
{R_{\mu}}^{\nu}= 2\left(F_{\mu \lambda} F^{\nu
\lambda}-\frac{1}{6}\,{\delta_{\mu}}^{\nu}\,F_{\alpha \beta}
F^{\alpha \beta}\right) - \frac{4}{l^2}\,{\delta_{\mu}}^{\nu}\,,
\label{einmax}
\end{equation}
\begin{equation}
\nabla_{\nu}F^{\mu\nu}+\frac{1}{2\sqrt{3}\sqrt{-g}}\,\epsilon^{\mu \alpha\beta\lambda\tau} F_{\alpha\beta}F_{\lambda\tau}=0\,.
\label{maxw}
\end{equation}
The general black hole solution of \cite{cclp} to these equations  can be written in the form
\begin{eqnarray}
\label{gsugrabh}
ds^2 & = & - \left( dt - \frac{a
\sin^2\theta}{\Xi_a}\,d\phi - \frac{b
\cos^2\theta}{\Xi_b}\,d\psi \right)\nonumber
\left[f \left( dt - \frac{a
\sin^2\theta}{\Xi_a}\,d\phi\, - \frac{b
\cos^2\theta}{\Xi_b}\,d\psi \right)\nonumber
\right. \\[2mm]  & & \left. \nonumber
+ \frac{2 Q}{\Sigma}\left(\frac{b
\sin^2\theta}{\Xi_a}\,d\phi + \frac{a
\cos^2\theta}{\Xi_b}\,d\psi \right) \right]
+ \,\Sigma
\left(\frac{r^2 dr^2}{\Delta_r} + \frac{d\theta^{\,2}}{
~\Delta_{\theta}}\right) \\[2mm] &&
+ \,\frac{\Delta_{\theta}\sin^2\theta}{\Sigma} \left(a\, dt -
\frac{r^2+a^2}{\Xi_a} \,d\phi \right)^2
+\,\frac{\Delta_{\theta}\cos^2\theta}{\Sigma} \left(b\, dt -
\frac{r^2+b^2}{\Xi_b} \,d\psi \right)^2 \nonumber
\\[2mm] &&
+\,\frac{1+r^2\,l^{-2}}{r^2 \Sigma } \left( a  b \,dt - \frac{b
(r^2+a^2) \sin^2\theta}{\Xi_a}\,d\phi
- \, \frac{a (r^2+b^2)
\cos^2\theta}{\Xi_b}\,d\psi \right)^2,
\end{eqnarray}
where
\begin{eqnarray}
f& =&\frac{\Delta_r - 2 a b\, Q -Q^2}{r^2 \Sigma}+ \frac{Q^2} {\Sigma^2}\,\,,~~~~ \Xi_a=1 - \frac{a^2}{l^2}\,\,,~~~~ \Xi_b=1 - \frac{b^2}{l^2}\,\,,
\nonumber \\[6mm]
\Delta_r &= &\left(r^2 + a^2\right)\left(r^2 +
b^2\right)\left(1+r^2l^{-2} \right)+ 2 a b\, Q + Q^2 - 2 M r^2\,, \nonumber \\[4mm]
\Delta_\theta & = & 1 -\frac{a^2}{l^2} \,\cos^2\theta
-\frac{b^2}{l^2} \,\sin^2\theta \,,~~~~~
\Sigma  =  r^2+ a^2 \cos^2\theta + b^2 \sin^2\theta \,
\label{gsugrametfunc}
\end{eqnarray}
We see that the metric is characterized  by the parameters of mass $M $, electric charge $ Q $ as well as by  two independent rotation parameters $ a $ and $ b $. The cosmological constant is taken to be negative determining the cosmological length scale  as  $ l^2= - 6/\Lambda $. Throughout this paper we suppose that the rotation parameters satisfy the relation $ a^2, b^2 < l^2 $.

For the potential one-form of the electromagnetic field, we have
\begin{equation}
A= -\frac{\sqrt{3}\,Q}{2 \,\Sigma}\,\left(dt- \frac{a
\sin^2\theta}{\Xi_a}\,d\phi -\frac{b \cos^2\theta}{\Xi_b}\,d\psi
\right)\,.\label{sugrapotform1}
\end{equation}

We recall that in their canonical forms, the Kerr-Newman-AdS metric in four dimensions as well as the Kerr-AdS metric in five dimensions are given in the Boyer-Lindquist  coordinates which are rotating at spatial  infinity. In order to be able to make an easy comparison  of our description with those in four and five dimensions, we  give the metric  (\ref{gsugrabh}) in the asymptotically rotating  Boyer-Lindquist coordinates $ x^{\mu}=\{t, r, \theta, \phi, \psi\}$  with $\mu=0, 1, 2, 3, 4 $ (see Ref.\cite{aliev1}). It is easy to see
that for  $ Q=0 $, it  recovers the five-dimensional Kerr-AdS solution of \cite{hhtr}.

The authors of \cite{cclp} have calculated the physical parameters and examined  the global structure and the supersymmetric properties of the solution in (\ref{gsugrabh}). In particular, they showed that for appropriate ranges of the parameters,  the solution is free of closed timelike curves(CTCs) and naked singularities,  describing a regular rotating charged black hole.

The determinant of the metric (\ref{gsugrabh}) does not involve the electric charge parameter $ Q $ and is given by
\begin{equation}
\sqrt{-g}= \frac{r \Sigma \sin\theta\,\cos\theta}{\Xi_a\,
\Xi_b}\,\,,\label{5ddeterminant}
\end{equation}
whereas, the contravariant metric components have the form
\begin{eqnarray}
g^{00}&=&-\frac{1}{\Sigma}\left\{
\frac{(r^2+a^2)(r^2+b^2)\left[ r^2+l^2(1-\Xi_a\Xi_b)\right]\,+2\,ab\left[ (r^2+a^2+b^2)\, Q+ a b M\right]}{\Delta_r} \nonumber \right. \\[2mm]  & & \left.
\nonumber
-l^2\left(1-\frac{\Xi_a\Xi_b}{\Delta_\theta}\right)\right\}\,,~~~~~
g^{11}=\frac{\Delta_r}{r^2\Sigma}\,\,,~~~~~~g^{22}=\frac{\Delta_\theta}{\Sigma}\,\,,
\\[3mm]
g^{03}&=&\frac{\Xi_a}{\Sigma} \left\{ \frac{a\, \Xi_b}{\Delta_\theta} - \frac{(r^2+b^2) \left[ b Q+ a\, \Xi_b  (r^2+a^2) \right]+ 2\, a b (a Q +b  M) }{\Delta_r} \right\}\,,\nonumber
\\[3mm]
g^{04}&=& \frac{\Xi_b}{\Sigma} \left\{ \frac{b\, \Xi_a}{\Delta_\theta} - \frac{(r^2+a^2) \left[ a Q+ b\, \Xi_a  (r^2+b^2) \right]+ 2\, a b (b Q +a  M) }{\Delta_r} \right\}\,,\nonumber
\\[3mm]
g^{33}&=&\frac{\Xi_a^2}{\Sigma}\left\{ \frac{\cot^2\theta+\Xi_b}{\Delta_\theta} + \frac{(r^2+b^2)\left[ b^2-a^2+(r^2+a^2)(1-\Xi_b) \right] -2b\,(a Q+b M)}{\Delta_r} \right\}\,,\nonumber
\\[3mm]
g^{44}&=&\frac{\Xi_b^2}{\Sigma}\left\{ \frac{\tan^2\theta+\Xi_a}{\Delta_\theta} + \frac{(r^2+a^2)\left[ a^2-b^2+(r^2+b^2)(1-\Xi_a) \right] -2a\,(b Q+a M)}{\Delta_r} \right\}\,,\nonumber
\\[3mm]
g^{34}&=& -\frac{\Xi_a \Xi_b}{\Sigma} \left\{ \phantom{\bigg\{ }\frac{a b}{l^2} \left[ \frac{1}{\Delta_\theta}- \frac{(r^2+a^2)(r^2+b^2)}{\Delta_r}  \right] + \frac{2 a  b M+ (a^2+b^2)\,Q}{\Delta_r}\phantom{\bigg\} }  \right\}\,.
\label{contras}
\end{eqnarray}

The horizons of the black hole are governed by the equation $ \Delta_r=0 $, which can be regarded as a  cubic equation with respect to $ r^2 $. It has two real  roots; $ r_1=r_{+}^2 $ and  $ r_2=r_{0}^2$. The largest of these roots, $ r_{+}^2 > r_{0}^2 $, represents the radius of the event horizon. However, when the equations
\begin{eqnarray}
\Delta_r&=& 0 \,\,, ~~~~~~\frac{d \Delta_r}{dr}=0
\label{extremeeq}
\end{eqnarray}
are satisfied simultaneously, the two roots coincide, $ r_{+}^2 = r_{0}^2 = r_{e}^2 $, representing the event horizon of an extreme black hole. From these equations, it follows  that the parameters of the extreme black hole must obey the relations
\begin{eqnarray}
2 M_{e}l^2 &=&  2 \left(r_{e}^2+ a^2 +b^2 +l^2 \right)r_{e}^2 +r_{e}^4 + a^2 b^2 + \left(a^2 + b^2\right) l^2\,,\\[2mm]
Q_{e}&= &\frac{r_{e}^2}{l} \left(2 r_{e}^2+ a^2 + b^2 +l^2\right)^{1/2} - a b\,.
\label{extparam}
\end{eqnarray}

The time translational and rotational (bi-azimuthal) isometries of the spacetime (\ref{gsugrabh})  are  defined by the Killing vector fields
\begin{equation}
{\bf \xi}_{(t)}= \partial / \partial t\,, ~~~~
{\bf \xi}_{(\phi)}= \partial / \partial \phi \, , ~~~~
{\bf \xi}_{(\psi)}= \partial / \partial \psi \,.
\label{3killings}
\end{equation}
Using these Killing vectors one can also introduce a corotating Killing vector
\begin{equation}
\chi = {\bf \xi}_{(t)}+ \Omega_{a}\,{\bf \xi}_{(\phi)}+
\Omega_{b} \,{\bf \xi}_{(\psi)}\,,\label{corotating}
\end{equation}
where $ \Omega_{a} $  and $ \Omega_{b} $ are the angular velocities of the event horizon in two independent orthogonal 2-planes of rotation. We have
\begin{eqnarray}
 \Omega_{a}&=&\frac{\Xi_a [a(r_+^2+b^2)+b Q]}{(r_+^2+a^2)(r_+^2+b^2)+a b Q }\,\,,~~~~~~~~~ \Omega_{b}=\frac{\Xi_b [b(r_+^2+a^2)+a Q] }{(r_+^2+a^2)(r_+^2+b^2)+a b Q}\,\,.
\end{eqnarray}
It is straightforward to show that the corotating Killing vector in (\ref {corotating}) is null on the event horizon of the black hole, i.e. it is tangent to the null generators of the horizon, confirming that the quantities  $ \Omega_{a} $  and $ \Omega_{b} $  are indeed the angular velocities of the horizon.

We also need  the electrostatic potential of the horizon relative to an infinitely distant point. It is given by
\begin{eqnarray}
\Phi_H & = - & A \cdot \chi   = -
\left(A_0 + \Omega_{a}\,A_{\phi} +\Omega_{b}\,A_{\psi}\right)|_{r=r_+ } \,\,.
\label{hpot}
\end{eqnarray}
Substituting into this expression the components of the potential in (\ref{sugrapotform1}), we find the explicit form for the electrostatic potential
\begin{equation}
\Phi_H=\frac{\sqrt{3}}{2}\,\frac{ Q r_+^2}{(r_+^2+a^2)(r_+^2+b^2)+a b Q }\,.
\end{equation}
It is also important to note that, in addition to the global isometries, the spacetime (\ref{gsugrabh})  also possesses hidden symmetries  generated by a second-rank Killing tensor. The existence of the Killing tensor ensures the complete separability of variables in the Hamilton-Jacobi equation for geodesic motion of uncharged particles \cite{dkl}. Below, we describe the separation of variables
in the Hamilton-Jacobi equation for charged particles.

\subsection{The Hamilton-Jacobi equation for charged particles}

The Hamilton-Jacobi equation  for a particle of electric charge $ e $ moving in the spacetime  under consideration  is given by
\begin{equation}
 \frac{\partial S}{\partial \lambda}+\frac{1}{2}g^{\mu\nu}\left(\frac{\partial S}{\partial x^\mu}-eA_\mu\right)\left(\frac{\partial S}{\partial x^\nu}-e A_\nu\right)=0 \,,
 \label{HJeq}
\end{equation}
where, $ \lambda  $ is  an  affine parameter.  Since the potential one-form  (\ref{sugrapotform1}) respects the Killing isometries (\ref{3killings}) of the spacetime as well,  $  \pounds_{\xi} A^{\mu}=0 $, we assume  that the action  $ S $  can be written in the form
\begin{equation}
S=\frac{1}{2}m^2\lambda - E t+L_\phi  \phi +L_\psi  \psi + S_r(r)+S_\theta(\theta)\,,
\label{ss}
\end{equation}
where the constants of motion represent  the mass  $ m  $,  the total energy $  E $ and the angular momenta  $ L_{\phi} $ and  $ L_{\psi} $ associated with the rotations in $ \phi $ and  $ \psi $ 2-planes. Substituting this action into equation (\ref{HJeq}) and using the  metric components in (\ref{contras}) along with the contravariant components of the potential
\begin{eqnarray}
A^{0}&=&\frac{\sqrt{3}\, Q}{2\Sigma }\,\, \frac{(r^2+a^2)(r^2+b^2)+a b Q}{\Delta_r}\,,\nonumber \\[2mm]
A^{3}&=& \frac{\sqrt{3}\, Q  \Xi_a}{2\Sigma }\,\,\,\frac{a (r^2+b^2)+b Q}{\Delta_r}\,, \nonumber \\[2mm]
A^{4}&=& \frac{\sqrt{3}\, Q  \Xi_b}{2\Sigma }\,\,\frac{b (r^2+a^2)+a Q}{\Delta_r}\,,
\label{potcontras}
\end{eqnarray}
and
\begin{equation}
A^\mu A_\mu=-\frac{3\,Q^2 r^2}{4\,\Sigma \Delta_r}\,,
\label{sqpot}
\end{equation}
we obtain two independent ordinary differential  equations for $ r $ and $ \theta$ motions:
\begin{eqnarray}
&& \frac{\Delta_r}{r^2} \left(\frac{dS_r}{dr}\right)^2 +\frac{\left(a b E-b \Xi_a  L_\phi-a \Xi_b  L_\psi\right)^2}{r^2} - \frac{\left[(r^2+a^2)(r^2+b^2)+a b Q \right]^2}{\Delta_r r^2} \cdot \nonumber
\\[2mm] &&
~~~~~ \left\{E- \frac{L_\phi\Xi_a \left[a(r^2+b^2)+b\,Q\right]}{(r^2+a^2)(r^2+b^2)+a b Q }-\frac{ L_\psi\Xi_b \left[b(r^2+a^2)+a Q \right]}{(r^2+a^2)(r^2+b^2)+a b Q}  \right.\nonumber \\[2mm] && \left.
~~~~~ - \frac{\sqrt{3}}{2}\,\frac{e Q r^2}{(r^2+a^2)(r^2+b^2)+a b Q }\right\}^2  +m^2 r^2 =-K\,,
\label{radeq1}
\end{eqnarray}

\begin{eqnarray}
&& \Delta_{\theta} \left(\frac{dS_\theta}{d\theta}\right)^2  +\frac{L_{\phi}^2 \,\Xi_{a}^2 \left(\cot^2\theta +\Xi_b\right) + L_{\psi}^2 \,\Xi_{b}^2 \left(\tan^2\theta +\Xi_a\right)- 2 a b l^{-2} L_{\phi} L_{\psi} \Xi_{a}\Xi_{b}}{\Delta_{\theta}}\nonumber \\[2mm] && ~~~+ \frac{ E^2 l^2 \left(\Delta_\theta - \Xi_a \Xi_b \right) - 2 E \left(a L_\phi + b L_\psi\right)\Xi_a \Xi_b }{\Delta_{\theta}}
+ m^2 \left(a^2 \cos^2\theta+ b^2\sin^2\theta\right)= K\,,
\label{angulareq1}
\end{eqnarray}
where $ K $ is a  constant of separation. The  complete separability in the Hamilton-Jacobi equation (\ref {HJeq}) occurs due to the existence of a new quadratic integral of motion $ K=K^{\mu\nu }p_{\mu} p_{\nu} \,$, which is associated with the hidden symmetries of the spacetime.  Here $ K^{\mu\nu }$ is an irreducible Killing tensor generating the hidden symmetries. Using equation (\ref{angulareq1}) along with  $ - m^2= g^{\mu\nu }p_{\mu} p_{\nu} \,$, we obtain  that  the Killing tensor is given by
\begin{eqnarray}
&&K^{\mu\nu}= \, l^2\left(1-\frac{\Xi_a\Xi_b}{\Delta_\theta}\right)\delta^\mu_t\delta^\nu_t + \frac{\Xi_a \Xi_b}{\Delta_\theta} \left[a\left( \delta^\mu_t \delta^\nu_\phi+\delta^\mu_\phi \delta^\nu_t\right)+b\left( \delta^\mu_t \delta^\nu_\psi+\delta^\mu_\psi \delta^\nu_t\right)\right] \nonumber \\[2mm] &&
+  \frac{1}{\Delta_\theta}\left[
\Xi_{a}^2 \left(\cot^2\theta +\Xi_b\right) \delta^\mu_\phi\delta^\nu_\phi + \Xi_{b}^2 \left(\tan^2\theta +\Xi_a\right) \delta^\mu_\psi\delta^\nu_\psi  -\frac{a b \Xi_a \Xi_b}{l^2} \left(\delta^\mu_\phi\delta^\nu_\psi+\delta^\mu_\psi\delta^\nu_\phi\right)\right]
\nonumber \\[2mm] &&
- \,g^{\mu\nu}\left(a^2\cos\theta^2 + b^2\sin\theta^2\right) + \Delta_\theta\, \delta^\mu_\theta\delta^\nu_\theta\,.
\label{killingt}
\end{eqnarray}
This expression agrees with that given in  \cite{dkl} up to terms involving symmetrized outer products of the Killing vectors. Similarly, for the vanishing cosmological constant, $ l \rightarrow \infty $, it
recovers the result of work \cite{fs1}.

\section{The Klein-Gordon equation}

We consider now the Klein-Gordon equation  for a scalar field with charge $ e $ and mass $ \mu $ in the background of the metric (\ref{gsugrabh}). It is given by
\begin{equation}
\left(D^{\mu}D_{\mu}-\mu^2 \right)\Phi= 0\,,
\label{KGeq}
\end{equation}
where $ D_{\mu}= \nabla_{\mu}- ie A_{\mu} $ and $ \nabla_{\mu} $  is a covariant derivative operator. Decomposing the indices as $ \mu= \{1,2 , M \} $ in which  $ M = 0,\, 3, \,4 $, we can write down the above equation in the form
\begin{eqnarray}
\frac{1}{r}\frac{\partial}{\partial r} \left(\frac{\Delta_r}{r}\, \frac{\partial \Phi}{\partial r}\right)
+\frac{1}{\sin 2\theta}
\frac{\partial}{\partial\theta}\left (\sin2 \theta \Delta_\theta \frac{\partial \Phi}{\partial \theta}
\right)+ \nonumber \\ [2mm]
\left(g^{MN}\frac{\partial^2 \Phi}{\partial x^M {\partial x^N}}
- 2 i e A^M \frac{\partial \Phi}{\partial x^N}-e^2 A_M A^M \right)\Sigma  &=& \mu^2 \Sigma \Phi \,.
\label{decomeq}
\end{eqnarray}
It is easy to show that with the components of $ g^{MN} $ given in (\ref{contras}) and with equations (\ref{potcontras}) and (\ref{sqpot}), this equation is manifestly separable in variables $ r $ and $ \theta $.  That is, one can assume  that its solution admits the ansatz
\begin{equation}
\Phi=e^{-i \omega t + i m_\phi  \phi +i m_\psi  \psi} S(\theta) R(r)\,,
\label{sansatz}
\end{equation}
where $ m_\phi $ and $ m_\psi $ are the ``magnetic" quantum numbers related to  $ \phi $ and  $ \psi $ 2-planes of rotation, so that they both must take integer values. In what follows, for the sake of certainty,  we restrict ourselves to the case of  positive frequency $ (\omega > 0) $ and positive $ m_\phi $ and $ m_\psi $ .

The substitution of  the expression (\ref{sansatz}) into equation (\ref{decomeq}) results in two decoupled ordinary differential equations for angular and radial functions. The angular equation is given by
\begin{eqnarray}
&&\frac{1}{\sin2 \theta}\frac{d}{d\theta}\left(\sin2\theta\Delta_\theta\frac{d S}{d\theta}\right) +\frac{1}{\Delta_\theta}\left[\omega^2 l^2 \left(\Xi_a \Xi_b - \Delta_\theta \right) -  m_\phi^2 \Xi_{a}^2 \left(\cot^2\theta +\Xi_b\right)-  m_\psi^2 \Xi_{b}^2 \left(\tan^2\theta +\Xi_a\right)\right.\nonumber \\[2mm] && \left.
~~~~~ +2 \,\Xi_a \Xi_b \left( \omega  a m_\phi + \omega  b m_\psi + \frac{a b} {l^{2}}\, m_\phi m_\psi \right) -  \Delta_\theta \mu^2 \left(a^2\cos^2\theta+b^2\sin^2\theta \right)\right] S = - \lambda S\,,
\label{angular1}
\end{eqnarray}
where $\lambda $ is a constant of separation. With regular boundary conditions at $ \theta=0 $ and  $ \theta=\pi/2 $, this equation  describes a well-defined Sturm-Liouville problem  with eigenvalues $ \lambda_{\ell}(\omega) $, where $\ell $ is thought of as an ``orbital" quantum number. The corresponding eigenfunctions are five-dimensional (AdS modified) spheroidal functions $ S(\theta)= S_{\ell\, m_\phi m_\psi}(\theta|a \omega\,, b\omega) $. In some special cases of interest, the eigenvalues were calculated in the Appendix , see equation (\ref{flat}).

The radial equation can be written in the form
\begin{equation}
\frac{\Delta_r}{r} \frac{d}{d r}\left(\frac{\Delta_r}{r} \frac{d R}{d r}\right) +U(r)\,R=0\,,
\label{radial1}
\end{equation}
where
\begin{eqnarray}
\label{rpotential}
&&U(r)=-\Delta_r \left[\lambda + \mu^2 r^2 + \frac{\left(a b\, \omega-b \Xi_a  m_\phi-a \Xi_b  m_\psi\right)^2}{r^2}\right]+\frac{\left[(r^2+a^2)(r^2+b^2)+ab\,Q \right]^2}{r^2} \cdot \nonumber
\\[4mm]
&&\left\{\omega- \frac{m_\phi\Xi_a \left[a(r^2+b^2)+b Q\right]}{(r^2+a^2)(r^2+b^2)+a b Q } - \frac{m_\psi\Xi_b \left[b(r^2+a^2)+a Q\right]}{(r^2+a^2)(r^2+b^2)+a b Q}
-\frac{\sqrt{3}}{2}\,\frac{ e Q r^2}{(r^2+a^2)(r^2+b^2)+a b Q }
\right\}^2.\nonumber \\[2mm]
\label{radpot1}
\end{eqnarray}
When the cosmological constant vanishes, $ l \rightarrow \infty $, the above expressions agree with those obtained in \cite{fs2} for a five-dimensional Myers-Perry black hole.

\section{The Superradiance Threshold}

The radial equation can be easily solved near the horizon. For this purpose, it is convenient to introduce a new radial function $ \mathcal{R}$ defined by
\begin{equation}
R= \left[\frac{r}{(r^2+a^2)(r^2+b^2)+a b Q}\right]^{1/2} \,\mathcal{R}\, \label{newrad}
\end{equation}
and a new radial, the so-called tortoise  coordinate  $ r_* $, obeying the relation
\begin{equation}
\frac{dr_*}{dr}=\frac{(r^2+a^2)(r^2+b^2)+a b Q}{\Delta_r}\,\,.
\label{tortoise}
\end{equation}
With these new definitions the radial equation (\ref{radial1}) can be transformed into the form
\begin{eqnarray}
\frac{d^2\mathcal{R}}{dr_*^2} +V(r)\mathcal{R}=0\,,
\label{radial2}
\end{eqnarray}
where the effective potential  is given by
\begin{eqnarray}
&&V(r)= -\frac{\Delta_r \left[r^2 \left(\lambda +\mu^2 r^2\right)+\left(a b\,\omega-b \Xi_a m_\phi-a \Xi_b m_\psi\right)^2\right]}{\left[(r^2+a^2)(r^2+b^2)+a b Q\right]^2}
-\frac{\Delta_r }{2 r u^{3/2}}\,\frac{d}{d r}\left(\frac{\Delta_r }{r u^{3/2}}\,\frac{d u}{d r}\right) + \nonumber
\\[4mm] &&
\left\{\omega- \frac{m_\phi\Xi_a \left[a(r^2+b^2)+b\,Q\right]}{(r^2+a^2)(r^2+b^2)+a b Q } - \frac{m_\psi\Xi_b \left[b(r^2+a^2)+a Q\right]}{(r^2+a^2)(r^2+b^2)+a b Q}
-\frac{\sqrt{3}}{2}\,\frac{e Q r^2}{(r^2+a^2)(r^2+b^2)+a b Q }\right\}^2\,. \nonumber\\[2mm]
\label{effective}
\end{eqnarray}
For brevity, we have also introduced
\begin{equation}
u=\frac{(r^2+a^2)(r^2+b^2)+a b Q}{r}\,.
\end{equation}
In what follows, we consider a massless scalar field, $ \mu=0 $.
We see that at the horizon $ r = r_+ $ $ (\Delta_r = 0) $, the effective potential in equation (\ref{effective})  becomes
\begin{eqnarray}
V(r_+ )&=&(\omega-m_\phi \Omega_{a}-m_\psi \Omega_{b}-e \Phi_{H})^2\,.
\label{p}
\end{eqnarray}
With this in mind, it is easy to verify that for an observer near the horizon, the asymptotic solution of the wave equation
\begin{equation}
\Phi=e^{-i \omega t + i m_\phi  \phi +i m_\psi  \psi}\, e^{-i(\omega- \omega_{p})r_*}S(\theta) \,,
\label{sansatzh}
\end{equation}
corresponds to an ingoing wave at the horizon. The threshold frequency
\begin{equation}
\omega_{p}= m_\phi \Omega_{a}+m_\psi \Omega_{b}+e \Phi_{H}
\label{bound}
\end{equation}
determines the frequency range
\begin{equation}
0< \omega < \omega_{p}
\label{fbound}
\end{equation}
for which, the phase velocity of the wave changes its sign. As in the four dimensional case \cite{press2}, this fact is the signature of the superradiance. That is, when the condition (\ref{fbound}) is fulfilled there must exist a superradiant outflow of energy from the black hole. From equation (\ref{bound}), it follows that the electric charge of the black hole changes the superradiance threshold frequency for charged particles.

Next, turning to the asymptotic behavior of the solution at spatial infinity, we recall  that in this region the AdS spacetime  reveals a box-like behavior. In other words, at spatial infinity the spacetime effectively acts as a reflective barrier. Therefore, we require the vanishing field boundary condition
\begin{equation}
\Phi\rightarrow 0 \,~~~~ as ~~~~  r\rightarrow \infty\, .
\label{boundinf}
\end{equation}
With the boundary conditions  (\ref{sansatzh}) and (\ref{boundinf}), namely, requiring a purely ingoing wave at the horizon and a purely damping wave at infinity, we arrive at a characteristic-value problem for complex frequencies of  quasinormal modes of the massless scalar field, see \cite{detweiler}. The imaginary part of these frequencies describes the damping of the modes.   A characteristic mode is stable if the imaginary part of its complex frequency is negative ({\it the positive damping}), while for the positive imaginary part, the mode  undergoes  exponential growth ({\it the negative damping}). In the latter case, the system will develop instability.

\section{Instability}

In this section we describe the instability  for small-size five-dimensional AdS black holes, $ r_+ \ll l $, in the regime of low-frequency perturbations. That is, we assume that the wavelength of the perturbations is much larger than the typical size of the horizon, $ 1/\omega \gg  r_+  $. In addition, we also assume slow rotation i.e. we restrict ourselves to linear order terms in rotation parameters $ a $ and $ b $. With these approximations, we can apply the similar method first developed by Starobinsky \cite{starobinsky1} and later on used
by many authors (see, \cite{cardoso2} and references therein) to construct the solutions of the radial equation (\ref{radial1}) in the region near the horizon and in the far-region. It is remarkable that there exists an intermediate region where the two solutions overlap and matching these solutions enables us to calculate the frequency of quasinormal modes and explore the (in)stability of these modes.

\subsection{Near-region solution}

For small and slowly rotating black holes, in the region close to the horizon, $ r- r_+ \ll 1/\omega $, and  in the regime of low-frequency perturbations, $ 1/\omega \gg  r_+ \, $, the radial equation (\ref{radial1}) can be approximated  by the equation
\begin{equation}
 4\Delta_x \frac{d}{dx} \left(\Delta_x \frac{dR}{dx}\right) +\left[ x_+^3 \left(\omega-\omega_p\right)^2- \ell(\ell+2)\Delta_x \right] R=0\,,
 \label{nearrad1}
 \end{equation}
where we have used the new radial  coordinate $ x=r^2 $ and
\begin{equation}
\Delta_x \simeq  x^2-2 M x +Q^2 = (x-x_+)(x-x_-)\,.
\end{equation}
The superradiance threshold frequency is given by equation (\ref{bound}), in which one must now take
\begin{eqnarray}
\Omega_a \simeq\frac{a}{x_+} +\frac{b Q}{x_+^2}\,\,,~~~~~~ \Omega_b \simeq \frac{b}{x_+} +\frac{a Q}{x_+^2}\,\,, ~~~~~~ \Phi_H \simeq
\frac{\sqrt{3}}{2}\frac{Q}{x_+}\,.
\label{nearvel}
\end{eqnarray}
In obtaining the above equations we have neglected the term involving $ r_+ ^2/l^2 $ as well as all terms with square and higher orders in rotation parameters. With the approximation employed,
the eigenvalues $ \lambda_{\ell} $ are replaced by their five-dimensional flat spacetime value $ \lambda_{\ell} \simeq \ell(\ell+2) $. (See the Appendix).

Next, it is convenient to define a new dimensionless variable
\begin{equation}
z=\frac{x-x_+}{x-x_-}\,\,,
\end{equation}
which in the near-horizon region goes to zero, $ z\rightarrow 0 $. Then equation (\ref{nearrad1}) can be put in the form
\begin{equation}
z (1-z) \frac{d^2R}{dz^2} +(1-z)\frac{d R}{dz} +\left[ \frac{1-z}{z}\,\Omega^2 -\frac{\ell(\ell+2)}{4(1-z)}\right]R=0\,,
\label{nearrad2}
\end{equation}
where
\begin{equation}
\Omega=\frac{x_+^{3/2}}{2}\, \frac{\omega-\omega_p}{x_+-x_-}\,.
\label{newsuperf}
\end{equation}
It is straightforward to check that the ansatz
\begin{equation}
R(z)=z^{i \Omega}\,(1-z)^{1+\ell/2}\,F(z)\,,
\end{equation}
when substituting into the above equation, takes us to the hypergeometric equation of the form  (\ref{hyperg1}) for the function $ F(z)= F(\alpha\,, \beta\,,\gamma, z) $ with
\begin{equation}
\alpha= 1+\ell/2 + 2 i \Omega\\,,~~~~~~~\beta= 1+\ell/2\,,~~~~~~~
\gamma=1+2i\Omega\,.
\label{nearradpara}
\end{equation}
The physical solution of this  equation  corresponding to the ingoing wave at the horizon, $ z\rightarrow 0 $, is given by
\begin{equation}
R(z)= A z^{-i \Omega}\,(1-z)^{1+\ell/2}\,F\left(1+\ell/2\,, 1+\ell/2 - 2 i \Omega\,, 1-2i\Omega\,, z\right)\,,
\label{nearphys}
\end{equation}
where $ A $ is a constant. For large enough values of the wavelength, this solution may overlap with the far-region solution. Therefore, we need to consider  the large $ r $  $  (z \rightarrow 1) $ limit of  this solution. For this purpose, we use  the functional relation between the hypergeometric functions of the arguments $ z $ and  $ 1- z $ \cite{abramowitz}, which in our case has the form
\begin{eqnarray}
&& F\left(1+\ell/2\,, 1+\ell/2 - 2 i \Omega\,, 1-2i\Omega\,, z\right)=\nonumber\\[2mm]&&
~~~~~~\frac{\Gamma(-1-\ell)\,\Gamma(1-2i\Omega) }{\Gamma(-\ell/2)\,\Gamma(-\ell/2-2i\Omega)}\,F\left(1+\ell/2\,, 1+\ell/2 - 2 i \Omega\,, 2+\ell\,, 1- z\right) + \nonumber\\[2mm]&&
~~~~~~\frac{\Gamma(1+\ell)\,\Gamma(1-2i\Omega) }{\Gamma(1+\ell/2)\,\Gamma(1+\ell/2-2i\Omega)}\,\left(1-z\right)^{-1-\ell}F\left(-\ell/2-2i\Omega\,,-\ell/2 \,, -\ell\,, 1- z\right).
\label{funtr1}
\end{eqnarray}
Taking  this into account in equation (\ref{nearphys}), we obtain that the large $ r $ behavior of the near-region solution is given by
\begin{eqnarray}
R \sim A  \Gamma(1-2i\Omega)\left[\frac{\Gamma(-1-\ell)\,(r_+^2- r_-^2)^{1+\ell/2}}{\Gamma(-\ell/2)\,\Gamma(-\ell/2-2i\Omega)}\,\,r^{-2-\ell} + \frac{\Gamma(1+\ell)\,(r_+^2- r_-^2)^{-\ell/2}}{\Gamma(1+\ell/2)\,\Gamma(1+\ell/2-2i\Omega)}\, \,r^{\ell}\right],
\label{larnear}
\end{eqnarray}
where we have also used the fact that $F(\alpha\,, \beta\,,\gamma, 0 )=1 $.

\subsection{Far-region solution}

In this region $ r_+\gg M $ the effects of the black hole are suppressed and the radial equation (\ref{radial1}) in this approximation is reduced to the form
\begin{equation}
\left(1+\frac{r^2}{l^2}\right) \frac{d^2R}{dr^2} + \left(\frac{3}{r}+\frac{5r}{l^2}\right)\frac{dR}{dr}
+\left[\frac{\omega^2}{1+\frac{r^2}{l^2}}-\frac{\ell(\ell+2)}{r^2} \right]R=0\,.
\end{equation}
Defining  a new variable
\begin{equation}
y= \left(1+\frac{r^2}{l^2}\right)\,,
\end{equation}
we can also put the equation into the form
\begin{equation}
y(1-y)\frac{d^2R}{dy^2} +\left(1-3y \right) \frac{dR}{dy} -\frac{1}{4}\,\left[\frac{\omega^2 l^2}{y}-\frac{\ell(\ell+2)}{y-1}\right]R=0\,.
\label{adsrad1}
\end{equation}
We note that this is an equation in a pure AdS spacetime and therefore,  we look for its solution satisfying the boundary conditions at infinity,  $ y\rightarrow\infty $,  and at the origin of the AdS space, $ y\rightarrow 1 $.

Again, one can show that the ansatz
\begin{equation}
R=y^{\omega l/2}(1-y)^{\ell/2}\, F(y)\,,
\end{equation}
transforms equation (\ref{adsrad1}) into the hypergeometric equation of the form (\ref{hyperg1}), where the parameters of the hypergeometric function  $ F(\alpha\,, \beta\,,\gamma\,, y ) $ are given by
\begin{equation}
\alpha= 2+\ell/2 + \omega l/2\,,~~~~~~\beta= \ell/2 + \omega l/2\,,~~~~~~\gamma =1+ \omega l\,.
\end{equation}

The solution of this equation vanishing at $ y \rightarrow \infty $, i.e. obeying the boundary condition (\ref{boundinf}) is given by
\begin{equation}
R(y)= B y^{-2 -\ell/2}\,(1-y)^{\ell/2}\,F\left(2+\ell/2+ \omega l/2\,\,,2+\ell/2- \omega l/2\,\,, 3 \,\,, 1/y\right)\,,
\label{nearphys1}
\end{equation}
where $  B $ is a constant. We are also interested in knowing  the small $ r  $  $ (y\rightarrow 1) $  behavior of this solution.  Using
the expansion of the hypergeometric function in (\ref{nearphys1}) in terms of the hypergeometric functions of the argument $ 1-y $  given by
\begin{eqnarray}
&& F\left(2+\ell/2+ \omega l/2\,\,,2+\ell/2- \omega l/2\,\,, 3 \,\,, 1/y\right)=\nonumber\\[3mm]&&
\frac{\Gamma(3)\Gamma(1+\ell)\,\,y^{2+\ell/2- \omega l/2}\,(y-1)^{-1-\ell}}{\Gamma(2+\ell/2+ \omega l/2 )\,\Gamma(2+\ell/2 - \omega l/2)}\,F\left(1-\ell/2-\omega l/2\,\,, -1-\ell/2 - \omega l/2\,\,, -\ell\,\,, 1- y\right)  \nonumber\\[3mm]&&
+ \frac{\Gamma(3)\Gamma(-1-\ell)\,\,y^{2+\ell/2+ \omega l/2}}{\Gamma(1-\ell/2+ \omega l/2 )\,\Gamma(1-\ell/2 - \omega l/2)}\,F\left(2+\ell/2+\omega l/2\,\,, \ell/2 + \omega l/2\,\,, 2+\ell\,\,, 1- y\right)\,,\nonumber\\
\label{funtr2}
\end{eqnarray}
we find  that for small values of $ r  $ the asymptotic  solution has the form
\begin{eqnarray}
&&R(r)\sim B \Gamma(3)(-1)^{\ell/2}\left[\frac{\Gamma(1+\ell)\,l^{2+\ell} \,\,r^{-2-\ell} }{\Gamma(2+\ell/2+\omega l/2)\,\Gamma(2+\ell/2-\omega l/2)}\right. \nonumber\\[3mm]
&&
\left.
~~~~~~~~~~~~~~~~~~~~~~~~~~~~~~~~~~~
+\,\frac{\Gamma(-1-\ell)\,l^{-\ell}\,\,r^{\ell}}{\Gamma(1-\ell/2+\omega l/2)\,\Gamma(1-\ell/2-\omega l/2)}\right]\,.
\label{farradsmall}
\end{eqnarray}
Requiring the regularity of this solution at the origin of the AdS space $(r=0)$, we obtain  the quantization condition
\begin{equation}
2+\ell/2-\omega l/2 = -n \,,
\label{quantization}
\end{equation}
where  $ n $ is a non-negative integer being a ``principal" quantum number. We recall that with this condition the gamma function $ \Gamma(2+\ell/2-\omega l/2) =\infty $.
Thus, we find that the discrete frequency spectrum for scalar perturbations in the five-dimensional AdS spacetime is given by
\begin{equation}
\omega_n=\frac{2n+\ell+4}{l}\,.
\label{fspectrum}
\end{equation}
This formula generalizes the four-dimensional result of works in \cite{cklemos, burgess} to five dimensions. Since at infinity the causal structure of the AdS black hole is similar to that of the pure AdS  background, it is natural to assume that equation (\ref{fspectrum}) equally well governs the frequency spectrum at large distances from the black hole. However, the important difference is related to the inner boundaries which are different; for the AdS spacetime  we have  $ r=0 $, while for the black  hole in this spacetime we have $ r=r_+ $. Therefore, to catch the effect of the black hole, the  solution (\ref{farradsmall}) must ``respond" to the ingoing wave condition at the boundary $ r=r_+ $. Physically, this means that one must take into account the possibility for tunneling of the wave through the potential barrier into the black hole and scattering back. As we have described  above, this would result in the quasinormal spectrum  with the complex frequencies
\begin{equation}
\omega= \omega_n + i \sigma\,,
\label{complexf}
\end{equation}
where $ \sigma $ is supposed to be a small quantity, describing the damping of the quasinormal modes. Taking this into account in equation (\ref{farradsmall}), we first note that \begin{equation}
\Gamma(2+\ell/2+\omega l/2)\,\Gamma(2+\ell/2-\omega l/2)= \Gamma(4+\ell+n+il\sigma/2)\,\Gamma(-n-il\sigma/2)\,.
\label{product}
\end{equation}
Next, applying to this expression the functional relations for the gamma functions \cite{abramowitz}
\begin{equation}
\Gamma(k+z)=(k-1+z)(k-2+z)\ldots(1+z)\Gamma(1+z)\,,~~~~~ \Gamma(z)\Gamma(1-z)=\frac{\pi}{\sin{\pi z}}\,,
\label{gammakpz}
\end{equation}
where $ k $ is a non-negative integer, it is easy to show that for $ l\sigma \ll 1 $
\begin{equation}
\Gamma(2+\ell/2+\omega l/2)\,\Gamma(2+\ell/2-\omega l/2)
=-\frac{2 i}{l \sigma}\,\frac{(\ell+3+n)!}{(-1)^{n+1}n!}\,.
 \label{product1}
\end{equation}
Similarly, one can also show that
\begin{eqnarray}
\Gamma(1-\ell/2+\omega l/2)\,\Gamma(1-\ell/2-\omega l/2)=\Gamma(-1-\ell-n)\,\Gamma(3+n)\,.
 \label{product2}
\end{eqnarray}
Substituting now these expressions into equation (\ref{farradsmall}), we obtain the desired form  of the far-region solution at small values of $ r $. It is given by
\begin{eqnarray}
R&=& B \Gamma(3)(-1)^{\ell/2}\left[
\frac{\Gamma(-1-\ell)\, l^{-\ell}\,\,r^{\ell}}
{\Gamma(-1-\ell-n)\,\Gamma(3+n)}+ i \sigma\,
\frac{(-1)^{n+1} n!\,\Gamma(1+\ell)}{2(3+\ell+n)!}\,\,l^{3+\ell}\, r^{-2-\ell} \right].
\label{farnearsol}
\end{eqnarray}

\subsection{Overlapping}

Comparing the  large $ r $ behavior of the near-region solution in (\ref{larnear}) with the small  $ r $ behavior of  the far-region solution in (\ref{farnearsol}), we conclude that there exists an intermediate  region  $ r_+ \ll r-r_+\ll 1/\omega $ where these solutions overlap. In this region  we can  match them which allows us to obtain the damping factor in the form
\begin{eqnarray}
\sigma &=&- 2 i \,\frac{(r^2_+- r^2_-)^{1+\ell}}{l^{3+2\ell}}
\frac{(3+\ell+n)!\,(1+\ell+n)! }{(-1)^{\ell}\, n! \,(2+n)! \,[\ell!(1+\ell)!]^2} \, \frac{\Gamma(1+\ell/2)\,\Gamma(1+\ell/2-2 i \Omega)}
{\Gamma(-\ell/2)\,\Gamma(-\ell/2-2 i \Omega )}\,\,.
\label{sigma}
\end{eqnarray}
We note that in this expression  the quantity $ \Omega $ is given by
\begin{equation}
\Omega=\frac{r_+^{3}}{2}\, \frac{\omega_n-\omega_p}{r_+^2-r_-^2}\,.
\label{newsuperf}
\end{equation}
It is also important to note that, in contrast to the related expression in four dimensions \cite{cardoso2},  equation (\ref{sigma}) involves the term $\ell/2 $ in the arguments of the gamma functions. Therefore, its further evaluation requires us to consider the cases of even and  odd  values  of $\ell $ separately.

\subsubsection{Even $ \ell $ }

In this case using the functional relations \cite{abramowitz}
 \begin{eqnarray}
&&\Gamma(k+ i z)\,\Gamma(k - i z)=\Gamma(1+ i z)\,\Gamma(1- i z)\prod_{j=1}^{k-1}\left(j^2+z^2\right)\,\,,\\[2mm]
&&\Gamma(1+ i z)\,\Gamma(1- i z)=\frac{\pi z}{\sinh \pi z}\,,
\label{func3}
\end{eqnarray}
one can show that
\begin{eqnarray}
\frac{\Gamma(1+\ell/2)\,\Gamma(1+\ell/2-2 i \Omega)}
{\Gamma(-\ell/2)\,\Gamma(-\ell/2-2 i \Omega )}=-2 i \Omega \,[(\ell/2)!]^2 \prod_{j=1}^{\ell/2}\left(j^2+4 \Omega^2 \right)\,.
\end{eqnarray}
Substituting this expression into equation (\ref{sigma}) we find that
\begin{eqnarray}
\sigma & = & -  \left(\omega_n-\omega_p\right) \, \frac{2 (3+\ell+n)!\,(1+\ell+n)! }{(-1)^{\ell}\, n! \,(2+n)! \,[\ell!(1+\ell)!]^2} \, \frac{r_+^3\,(r^2_+- r^2_-)^{\ell}}{l^{3+2\ell}}
[(\ell/2)!]^2 \prod_{j=1}^{\ell/2}\left(j^2+4 \Omega^2 \right)\,.\nonumber\\
\label{sigmaf}
\end{eqnarray}
We see that the sign of this expression crucially depends on the sign of the factor $ \left(\omega_n-\omega_p\right)$   and in the superradiant regime $ \omega_n < \omega_p $  it is positive. In other words, we have the negative damping effect, as  we have discussed at the end of Sec.IV, resulting in exponential growth of the modes with characteristic time scale $ \tau=1/\sigma $. Thus, the small AdS black holes under consideration become unstable to  the superradiant scattering of massless scalar perturbations of even $ \ell $ or equivalently of even sum  $ m_\phi + m_\psi  $. We recall that we consider the positive frequency modes and the positive magnetic quantum numbers $ m_\phi $  and $ m_\psi $\,.

\subsubsection{Odd $ \ell $ }

In order to evaluate the combination of the gamma functions appearing in equation (\ref{sigma}) for odd values of  $ \ell $, we appeal to the relations \cite{abramowitz}
\begin{eqnarray}
\Gamma\left(k+\frac{1}{2}\right)=\pi^{1/2}2^{-k}(2k-1)!!\,,~~~~~
\Gamma\left(\frac{1}{2}+iz\right)\Gamma\left(\frac{1}{2}-iz\right)=\frac{\pi}{\cosh{\pi z}}\,.
\label{funct}
\end{eqnarray}
Using these relation along with those given in (\ref{gammakpz}), after some algebra, we obtain that
\begin{eqnarray}
 \frac{\Gamma(1+\ell/2)\,\Gamma(1+\ell/2-2 i \Omega)}
{\Gamma(-\ell/2)\,\Gamma(-\ell/2-2 i \Omega )}
 =
 \frac{(\ell!!)^2 }{ 2^{1+\ell}}\,\prod_{j=1}^{(\ell+1)/2} \left[\left(j-\frac{1}{2}\right)^2+4 \Omega^2\right].
\end{eqnarray}
With this in mind, we put  equation (\ref{sigma}) in the form
\begin{eqnarray}
\sigma &=&- i \,\frac{(r^2_+- r^2_-)^{1+\ell}}{l^{3+2\ell}}
\frac{(3+\ell+n)!\,(1+\ell+n)!\,(\ell!!)^2 }{(-1)^{\ell}\,2^{\ell}\, n! \,(2+n)! \,[\ell!(1+\ell)!]^2} \,\,\prod_{j=1}^{(\ell+1)/2} \left[\left(j-\frac{1}{2}\right)^2+4 \Omega^2\right].
\label{sigmaoddf}
\end{eqnarray}
We see that this expression is purely imaginary and it does not change the sign in the superradiant regime. In other words, these modes do not undergo any damping, but they do oscillate with frequency-shifts.

\section{Conclusion}

In this paper, we have discussed the instability properties of small-size,  $ r_+\ll l,$  charged AdS black holes with two rotation parameters, which are described by the  solution of  minimal five-dimensional  gauged supergravity recently found in \cite{cclp}. The remarkable symmetries  of this  solution allow us to perform  a complete  separation of variables in the field equations governing  scalar perturbations in the background of the AdS black holes.

We have begun with demonstrating the separability of variables in the Hamilton-Jacobi equation for massive charged particles as well as in the Klein-Gordon equation for a massive charged scalar field. In both cases, we have presented the decoupled radial and angular equations in their most compact form. Next, exploring the behavior of the radial equation near the horizon, we have found the threshold frequency for the superradiance of these black holes. Restricting ourselves to slow rotation  and to low-frequency perturbations, when the characteristic wavelength scale is much larger than the typical size of the black hole, we have constructed the solutions of  the radial equation in the region close to the horizon  and in the far-region of the spacetime. Performing the matching of these solutions in an overlapping region of their validity, we have derived  an analytical formula for the frequency spectrum of the quasinormal modes.

Analyzing  the imaginary part of the spectrum for modes of even  and odd $ \ell $ separately, we have revealed a new feature: In the regime of superradiance only the modes of even $ \ell $ undergo the negative damping, exponentially growing their amplitudes. On the other hand,  the modes of odd $ \ell $ turn out to be not sensitive to the regime of superradiance, oscillating without any damping, but with frequency-shifts. This new feature is inherent in the five-dimensional AdS black hole spacetime and absent in four dimensions where the small-size AdS black holes exhibit the instability to all modes of scalar perturbations in the regime of superradiance \cite{cardoso2}.

We emphasize once again that our result  was obtained in the regime of low-frequency perturbations and for small-size, slowly rotating AdS black holes. Therefore, its validity is guaranteed for a certain range of the perturbation frequencies and parameters of the black holes within the approximation employed. The full analysis beyond this approximation requires a numerical  work. Meanwhile, one should remember that the characteristic oscillating modes for the instability to occur (for superradiance) are governed by the radius of the AdS space. This means that the instability will not occur for an arbitrary range of the black hole parameters. We also emphasize that the different stability properties of  even and odd modes of scalar perturbations arise only in the  five-dimensional case with reflective boundary conditions. The physical reason for this  is apparently related with  the ``fermionic constituents" of the five-dimensional AdS black hole. Therefore,  it would be interesting to explore this effect in the spirit of work \cite{maldacena} using  an effective string theory picture, where even $ \ell $ refers to bosons and odd $ \ell $ to fermions. This is a challenging project for future work.

\section{Acknowledgments}

A. N. thanks the Scientific and Technological Research Council of Turkey (T{\"U}B\.{I}TAK) for partial support under the Research Project No. 105T437. O. D. also thanks T{\"U}B\.{I}TAK  for a postdoctoral fellowship through the Programme BIDEB-2218.

\appendix*

\section{Angular Equation}

The angular equation (\ref{angular1}) can be transformed into  a second-order Fuchsian equation by defining a new variable $ z=\sin^2\theta $. Performing straightforward calculations, we obtain \begin{eqnarray}
&& \frac{d^2 S}{dz^2}+\left( \frac{1}{z}+\frac{1}{z-1}+\frac{1}{z-c}\right) \frac{dS}{dz}
+\frac{1}{4 z(1-z)\Delta_z^2}\left[-\Delta_z \omega^2 l^2 - \frac{m_\phi^2 \Xi_a^2}{z} - \frac{m_\psi^2 \Xi_b^2}{1-z}
\right.\nonumber \\[2mm] && \left.
~~~~~ +\left(\Xi_a - \Xi_b\right)\left(m_\phi ^2\Xi_a-m_\psi  ^2\Xi_b \right) +\Xi_a \Xi_b l^2 \left(\omega+\frac{a}{l^2}m_\phi +\frac{b}{l^2}m_\psi  \right)^2
\right.\nonumber \\[2mm] && \left.
~~~~~ + \,\Delta_z\left(\lambda -\mu^2[ b^2 z + a^2(1-z)]\right)\right] S=0\,,
\label{fuchs}
\end{eqnarray}
where
\begin{eqnarray}
\Delta_z &=& \Xi_a+ (\Xi_b - \Xi_a) z\,,~~~~~~c=\frac{\Xi_a}{\Xi_a - \Xi_b}\,.
\label{zmfunc}
\end{eqnarray}
This equation has  four regular singular points $ z=0,\,1,\,c,\,\infty  $ and therefore can also be put in the form of the Heun equation \cite{Ronveaux}
\begin{equation}
\frac{d^2 H}{dz^2}+\left(\frac{1+2\alpha}{z}+\frac{1+2\beta}{z-1}+\frac{1+2\gamma}{z-c}\right)\frac{dH }{dz} +\frac{
\varepsilon \delta\, z+\eta}{z(z-1)(z-c)}\,H=0\,,
\label{heun}
\end{equation}
where the functions $ H(z) $ and  $ S (z) $ are related by
\begin{equation}
S(z)=z^{\alpha}(1-z)^{\beta}(z-c)^{\gamma}H(z)\,,
\end{equation}
and the associated parameters  are given by
\begin{eqnarray}
&&2 \alpha= m_\phi\,,~~~~~2\beta=m_\psi \,,~~~~~2 l \gamma= \omega l^2 +a m_\phi+b m_\psi\,,\nonumber\\
&& 2\delta= m_\phi+m_\psi + 2\left(\gamma+1 + \sqrt{1+l^2 \mu^2/4}\right)\,,\nonumber\\
&& 2\varepsilon= m_\phi+m_\psi + 2\left(\gamma+1 - \sqrt{1+l^2 \mu^2/4}\right)\,,\nonumber\\[1mm]
&&2\eta=-\left(2 \gamma+m_\phi\right)(1+m_\phi)- c (m_\phi+m_\psi+m_\phi m_\psi)\nonumber \\
&&\quad +\frac{c}{2\Xi_a}\left[\lambda- \mu^2 a^2 -\omega^2 l^2+\Xi_b(4 \gamma^2-m_\phi^2-m_\psi^2) \right]\,.
\end{eqnarray}
We note that the regularity of the point at $ z=\infty  $ gives the following relation between the parameters
\begin{equation}
\delta+\varepsilon= 2(\alpha+\beta+\gamma+1)\,.
\end{equation}
We also note that the regularity of the solutions requires that the magnetic quantum numbers $ m_\phi $ and  $ m_\psi $  must take non-negative values. Therefore, below we imply only these values for $ m_\phi $ and  $ m_\psi $. It turns out that for some special cases, namely, when the black hole has two equal rotation parameters $ (a = b, ~\Xi_a=\Xi_b=\Xi) $ or its rotation is slow enough, the above equation has only three regular singular points $ z=0,\,1,\,\infty $. That is, the corresponding solution can be expressed in terms of the hypergeometric functions. We consider now these cases separately.

\subsection{Equal rotation parameters}

In this case equation (\ref{fuchs}) reduces to the form
\begin{eqnarray}
&&z(1-z)\frac{d^2 S}{dz^2} + \left(1-2 z\right)\frac{d S}{dz} +\frac{1}{4}\left[\left(\frac{\omega l^2 + a (m_\phi+m_\psi)}{l}\right)^2
\right.\nonumber \\[2mm] && \left.
~~~~~~~~~~ - \frac{m_\phi^2}{z} - \frac{m_\psi^2}{1-z} +\frac{\lambda - \mu^2 a^2 -\omega^2 l^2}{\Xi}\right] S=0\,.
\label{equalp}
\end{eqnarray}
One can easily verify that with the substitution
\begin{equation}\label{hypertrans}
S(z)=z^{m_\phi/2} (1-z)^{m_\psi/2} F(z)\,,
\end{equation}
equation (\ref{equalp}) goes over into the standard hypergeometric differential equation
\begin{eqnarray}
&& z(1-z)\frac{d^2 F}{dz^2} +\left[\gamma-(\alpha +\beta+1)\,z\right] \frac{d F}{dz} - \alpha \beta F = 0\,,
\label{hyperg1}
\end{eqnarray}
where the parameters are given by
\begin{eqnarray}
&& 2\alpha= 1 + \delta+ m_\phi +m_\psi\,,~~~~ 2 \beta= 1 - \delta + m_\phi +m_\psi\,, ~~~~ \gamma=1+ m_\phi\,,
\nonumber \\[3mm]
&&
\delta^2= 1+\left(\frac{\omega l^2 + a (m_\phi+m_\psi)}{l}\right)^2
+\frac{\lambda - \mu^2 a^2 -\omega^2 l^2}{\Xi}\,.
\label{hyoerg1coef}
\end{eqnarray}
The general solution of this equation for $ z\in(0,1) $  has the form (see, for instance \cite{abramowitz})
\begin{eqnarray}
F(z)=A_1 \,z^{1-\gamma}\, F(\alpha-\gamma+1,\,\beta-\gamma+1\,,2-\gamma ,\,z)+ B_1\, F(\alpha,\,\beta,\,\gamma,\,z)\,,
\label{hypgeo1}
\end{eqnarray}
where $ A_1 $ and $ B_1 $  are constants. From the regularity of the solution at  $ z=0 $ and $ z=1 $, we obtain $ A_1 =0 \,$ and
\begin{eqnarray}
\lambda &= &\Xi\left[(2j+m_\phi +m_\psi  )(2j+m_\phi +m_\psi  +2)- 2\omega a(m_\phi+m_\psi)-\frac{a^2 (m_\phi+m_\psi)^2}{l^2}\right] \nonumber\\
&+& a^2(\omega^2+ \mu^2)\,,
\end{eqnarray}
where $ j $ is a non-negative integer.
Introducing now $ \ell=2j+m_\phi +m_\psi \,$, we can put this expression in the form
\begin{eqnarray}
\lambda &=&\Xi\left[\ell (\ell+2)- 2\omega a(m_\phi+m_\psi)-\frac{a^2 (m_\phi+m_\psi)^2}{l^2}\right] + a^2(\omega^2+ \mu^2)\,.
\label{eigenv1}
\end{eqnarray}
We note that the  new integer $ \ell $  being the orbital quantum number must obey the condition  $ \ell \geq m_\phi +m_\psi \,$. Thus, it must take even (odd) values if the sum $ m_\phi +m_\psi \,$ is even (odd). We also note that in the $ a=b $ case, the eigenvalues are the same as in the absence of rotation (with redefinition of the separation constant). That is,
\begin{equation}
\lambda_{\ell}=\ell(\ell+2)\,.
\label{flat}
\end{equation}
This is in agreement with the result obtained in \cite{fs2}.

\subsection{Slow Rotation}

When the rotation of the black hole is slow enough, we can discard all terms of higher than linear order in rotation parameters $ a $ and  $ b $.  In this case the angular equation (\ref{fuchs}) takes the simple  form
\begin{equation}
z(1-z)\frac{d^2 S}{dz^2} + \left(1-2 z\right)\frac{d S}{dz}
-\frac{1}{4}\left(\frac{m_\phi^2 }{z} + \frac{m_\psi^2}{1-z}- \lambda_{\ell} \right) S=0\,,
\label{hyperg2}
\end{equation}
where
\begin{equation}
\lambda_{\ell}=\lambda + 2\omega (a m_\phi +b m_\psi)\,.
\end{equation}
The similar substitution as in (\ref{hypertrans}) transforms this equation into equation (\ref{hyperg1}) with the parameters
\begin{equation}
2\alpha= 1+\sqrt{1+\lambda_{\ell}} + m_\phi +m_\psi \,,~~~~ 2 \beta=  1-\sqrt{1-\lambda_{\ell}} + m_\phi +m_\psi \,,~~~~
\gamma=1+ m_\phi\,.
 \label{hyoerg2coef}
\end{equation}
It is easy to verify that its  regular solution at $ z=0 $ and $ z=1 $ implies the eigenvalues given in (\ref{flat}).

\end{document}